\documentclass[twocolumn]{aa}
\usepackage{graphicx}
\usepackage[intlimits,sumlimits]{amsmath}
\usepackage{times,epsfig} 
\usepackage{natbib}
\usepackage{amssymb}
\usepackage{mathrsfs}  
\bibpunct{(}{)}{;}{a}{}{,} 
\usepackage{epstopdf}

\usepackage{amsmath}
\usepackage[english]{babel}
\usepackage{longtable}
\usepackage{url}
\usepackage{lscape}
\usepackage{hyperref}

\begin{document}

\title{SN~2006oz: rise of a super-luminous supernova observed by the SDSS-II SN Survey}
\titlerunning{The super-luminous SN~2006oz}

\author{
G.~Leloudas\inst{\ref{inst1},\ref{inst14},\ref{inst15}}
\and E.~Chatzopoulos\inst{\ref{inst2}}
\and B.~Dilday\inst{\ref{inst3},\ref{inst4}}
\and J.~Gorosabel\inst{\ref{inst6}}
\and J.~Vinko\inst{\ref{inst2},\ref{inst7}}
\and A.~Gallazzi\inst{\ref{inst1}}
\and J.~C.~Wheeler\inst{\ref{inst2}}
\and B.~Bassett\inst{\ref{inst8},\ref{inst9},\ref{inst10}}
\and J.~A.~Fischer\inst{\ref{inst24}}
\and J.~A.~Frieman\inst{\ref{inst11},\ref{inst12},\ref{inst13}}
\and J.~P.~U.~Fynbo\inst{\ref{inst1}}
\and A.~Goobar\inst{\ref{inst14},\ref{inst15}}
\and M.~Jel\'inek\inst{\ref{inst6}}
\and D.~Malesani\inst{\ref{inst1}}
\and R.~C.~Nichol\inst{\ref{inst16}}
\and J.~Nordin\inst{\ref{inst22},\ref{inst23}}
\and L.~\"Ostman\inst{\ref{inst20}}
\and M.~Sako\inst{\ref{inst24}}
\and D.~P.~Schneider\inst{\ref{inst17},\ref{inst18}}
\and M.~Smith\inst{\ref{inst8}, \ref{inst21}}
\and J.~Sollerman\inst{\ref{inst14},\ref{inst19}}
\and M.~D.~Stritzinger\inst{\ref{inst14},\ref{inst19},\ref{inst99}}
\and C.~C.~Th\"one\inst{\ref{inst6}}
\and A.~de~Ugarte~Postigo\inst{\ref{inst1},\ref{inst6}}
}

\institute{
Dark Cosmology Centre, Niels Bohr Institute, University of Copenhagen, 2100 Copenhagen, Denmark \label{inst1}
\and Department of Astronomy, University of Texas at Austin, Austin, TX, USA  \label{inst2}
\and Las Cumbres Observatory Global Telescope Network, Goleta, CA 93117, USA  \label{inst3}
\and Department of Physics, University of California, Santa Barbara, Broida Hall, Santa Barbara, CA 93106-9530, USA  \label{inst4}
\and Instituto de Astrof\'isica de Andaluc\'ia (IAA-CSIC), Granada, Spain  \label{inst6}
\and Department of Optics and Quantum Electronics, University of Szeged, Szeged, Hungary  \label{inst7}
\and African Institute for Mathematical sciences, Muizenberg, Cape Town, South Africa  \label{inst8}
\and South African Astronomical Observatory, Cape Town, South Africa \label{inst9}
\and Mathematics Department, University of Cape Town, Cape Town, South Africa \label{inst10}
\and University of Pennsylvania, 209 South 33rd Street, PA 19104, USA \label{inst24}
\and Department of Astronomy and Astrophysics, The University of Chicago, Chicago, IL 60637, USA \label{inst11}
\and Kavli Institute for Cosmological Physics, The University of Chicago, Chicago, IL 60637, USA \label{inst12}
\and Center for Particle Astrophysics, Fermi National Accelerator Laboratory, Batavia, IL 60510, USA \label{inst13}
\and The Oskar Klein Centre, Stockholm University, Albanova University Centre, 10691 Stockholm, Sweden \label{inst14}
\and Department of Physics, Stockholm University, Albanova University Centre, 10691 Stockholm, Sweden \label{inst15}
\and Institute of Cosmology and Gravitation, Dennis Sciama Building, University of Portsmouth, Portsmouth, PO1 3FX, UK \label{inst16}
\and Space Sciences Laboratory, University of California Berkeley, Berkeley, CA 94720, USA  \label{inst22}
\and E. O. Lawrence Berkeley National Lab, 1 Cyclotron Rd., Berkeley, CA 94720, USA  \label{inst23}
\and Institut de F\'isica d'Altes Energies, Universitat Autonoma de Barcelona, 08193 Bellaterra (Barcelona), Spain \label{inst20}
\and Department of Astronomy and Astrophysics, The Pennsylvania State University, University Park, PA 16802, USA \label{inst17}
\and Institute for Gravitation and the Cosmos, The Pennsylvania State University, University Park, PA 16802, USA \label{inst18}
\and Astrophysics, Cosmology and Gravity Centre, Department of Mathematics and Applied Mathematics, University of Cape Town, Rondebosch 7700, South Africa\label{inst21}
\and Department of Astronomy, Stockholm University, Albanova University Centre, 10691 Stockholm, Sweden \label{inst19}
\and Department of Physics and Astronomy, Aarhus University, Ny Munkegade 120, 8000 Aarhus C, Denmark \label{inst99}
}

%\offprints{\email{giorgos@dark-cosmology.dk}}
\date{Received 22 November 2011 / Accepted 29 February 2012}

\abstract{A new class of super-luminous transients has recently been  identified. These objects reach absolute luminosities of $M_u$~$<$~$-$~$21$,  lack hydrogen in their spectra, and are exclusively discovered by non-targeted surveys because they are associated with very faint galaxies.}
{We aim to contribute to a better understanding of these objects by studying SN~2006oz, a newly-recognized member of this class.}
{We present multi-color light curves of SN~2006oz from the SDSS-II SN Survey that cover its rise time, as well as an optical spectrum that shows that the explosion occurred at $z \sim 0.376$.
We fitted black-body functions to estimate the temperature and radius evolution of the photosphere and used the parametrized code SYNOW to model the spectrum.
We constructed a bolometric light curve and compared it with explosion models. 
In addition, we conducted a deep search for the host galaxy with the 10 m GTC telescope.
} 
{
The very early light curves show a dip in the $g$- and $r$-bands and a possible initial cooling phase in the $u$-band before rising to maximum light. 
The bolometric light curve shows a precursor  plateau with a duration of 6-10 days in the rest-frame.
A lower limit of $M_u < - 21.5$ can be placed on the absolute peak luminosity of the SN, while the rise time is constrained to be at least 29 days.
During our observations, the emitting sphere doubled its radius to $\sim 2 \times 10^{15}$~cm, while the temperature remained hot at $\sim$15000~K. As for other similar SNe, the spectrum is best modeled with elements including \ion{O}{ii} and \ion{Mg}{ii}, while we tentatively suggest that \ion{Fe}{iii} might be present.
The host galaxy is detected in $gri$ with 25.74 $\pm$ 0.19,  24.43 $\pm$  0.06, and 24.14 $\pm$  0.12, respectively. It is a faint dwarf galaxy with $M_g = -16.9$.
}
{
We suggest  that the precursor plateau might be related to a recombination wave in a circumstellar medium (CSM) and discuss whether this is a common property of all similar explosions. 
The subsequent rise can be equally well described by input from a magnetar or by ejecta--CSM interaction, but the models are not well constrained owing to the lack of post-maximum observations, 
and CSM interaction has difficulties accounting for the precursor plateau self-consistently.
Radioactive decay is less likely to be the mechanism that powers the luminosity. 
The host is a moderately young and star-forming, but not a starburst, galaxy.
}

\keywords{supernovae: general, supernovae: individual: SN~2006oz}

\maketitle

\section{Introduction}

Historically, supernovae (SNe) have usually  been discovered by monitoring bright, nearby galaxies.
With a few exceptions, the supernovae discovered in this targeted way fit well within the traditional SN classification scheme  \citep[e.g.][]{1997ARA&A..35..309F}.
During the last few years, however, rolling searches, such as the Texas SN Search \citep{2005AAS...20717102Q} and the Catalina Real-Time Transient Survey \citep{2009ApJ...696..870D} have changed our view of stellar explosions. 
In particular, it has been shown that interesting transients were previously missed exactly because they occur in environments different from those probed by the traditional SN searches. Remarkably, these include the brightest SNe ever recorded  (hereafter super-luminous supernovae, SLSNe) with luminosities exceeding those of   SNe~Ia by %$>$ 
10--100 times.

Many of these SLSNe are associated with SNe~IIn and have high luminosities attributed (at least partly) to the interaction of the ejecta with a dense H-rich CSM \citep[e.g.][]{2007ApJ...659L..13O,2007ApJ...666.1116S,2008ApJ...686..467S}. 
SN~2005ap, in contrast, showed only weak evidence for  hydrogen and yet reached an absolute magnitude of $M_R<-$22.5 \citep{2007ApJ...668L..99Q}.
A seemingly unrelated object of peculiar nature was discovered by the Supernova Cosmology Project (SCP) with HST  \citep{2009ApJ...690.1358B}. 
SCP06F6 had unprecedented spectra and light curves, but the lack of a robust redshift estimate left this study inconclusive with respect to its nature 
\citep[see also][]{2009ApJ...704.1251C,2009ApJ...697L.129G,2010NewA...15..189S}.

Significant progress was made when \cite{2011Natur.474..487Q} showed that four objects detected by the Palomar Transient Factory \citep[PTF;][]{2009PASP..121.1334R,2009PASP..121.1395L} could be grouped together with SN~2005ap and  SCP06F6 to form a distinct class of H-poor SLSNe. 
The redshifts of these objects were identified by the detection of the \ion{Mg}{ii} $\lambda$2800 doublet and SCP06F6 was  shown to be at a much higher redshift  ($z=1.189$) than originally estimated.
All objects had similar spectra, blue colors, and relatively symmetric light curves. Their typical absolute magnitude is $M \sim -$21.5 and their explosion mechanism remains a mystery.
Possible suggestions  
include pulsational pair-instability \citep{2007Natur.450..390W},  the powering of the ejecta by a magnetar \citep{2010ApJ...717..245K,2010ApJ...719L.204W}, or interaction with an CSM \citep{2011ApJ...729L...6C,2010arXiv1009.4353B,2011arXiv1110.3807M}. 
A third class of SLSNe, also H-poor, but different from those in  \cite{2011Natur.474..487Q}, 
might be represented by SN~2007bi, which has been proposed 
\citep{2009Natur.462..624G} to be a pair-instability event, although this explanation is not unique \citep{2010A&A...512A..70Y,2010ApJ...717L..83M}.

One of  the objects studied by \cite{2011Natur.474..487Q}, SN~2010gx, was also extensively followed by \cite{2010ApJ...724L..16P}.
By obtaining spectra at later phases, they managed to demonstrate that this SLSN transitioned to a SN~Ic, showing that there is a possible link between these energetic explosions.
Recently, two more SNe belonging to this intriguing class were discovered by Pan-STARRS1 at $z \sim$ 0.90 \citep{2011ApJ...743..114C}.
These objects showed no signs of deceleration  in their expansion velocities during observations obtained over a period of about three weeks around maximum light.

A common characteristic for H-poor SLSNe is that they are systematically found in faint galaxies. Indeed, to date, only three of these hosts have been detected; there are only upper limits on the others (typically $M_B > -$18), suggesting that they are probably  metal-poor  \citep{2011ApJ...727...15N}. 
It has been speculated that low metallicity might be an indispensable ingredient to produce SLSNe.  

One reason that our understanding of these objects is limited is that 
the available data are sparse and  incomplete:
observations are often obtained in 1 or 2  neighboring filters, 
and when multi-color light curves are available, they cover only part of the SN evolution.
The study of SLSNe is a relatively new topic and any complementary dataset (especially covering the critical early phases) 
constitutes a valuable contribution to the field.

SN~2006oz was discovered by the SDSS-II SN Survey \citep{2008AJ....135..338F} toward the end of the 2006 observing season
at R.A. = $22^{\rm h}08^{\rm m}53\fs56$, Dec = $+00^{\circ}53^{\prime}50\farcs4$ (J2000). 
Labeled internally as a `strange hostless transient', 
it was initially classified \citep{2006CBET..762....3S} as a \textit{possible} SN Ib based on a spectrum obtained at the Nordic Optical Telescope (NOT). 
In the analysis by \cite{2011A&A...526A..28O}, it was given a SN~II designation, noting, however, that it matched less than five templates in SNID \citep{2007ApJ...666.1024B}.
Today, we know that these misclassifications were due to the lack of suitable comparison spectra in the literature, and, 
here, we identify it as an H-poor SLSN, as defined by \cite{2011Natur.474..487Q}. 
Our observations and results are presented in the next section. Section \ref{sec:disc} contains the discussion and Section  \ref{sec:conc} our concluding remarks.

\section{Observations and results}
\label{sec:obs}

\begin{figure*}
\begin{center}
\includegraphics[width=\textwidth]{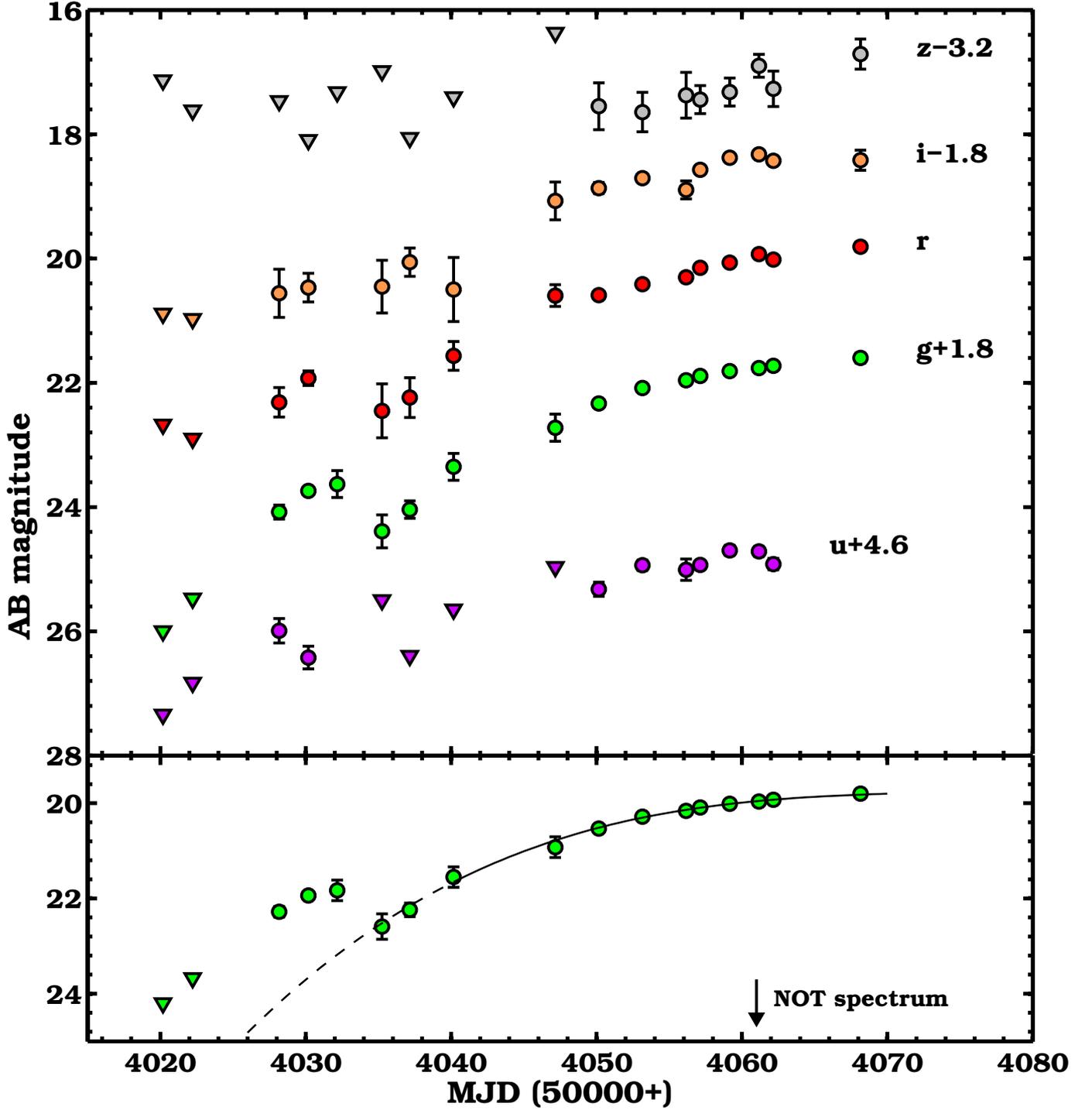}
\caption[]{$ugriz$ light curves of SN~2006oz obtained by the SDSS SN Survey, shifted as indicated for presentation purposes. Triangles denote 3$\sigma$ upper limits. 
The lower panel focuses on the $g$-band evolution shown together with a cubic polynomial fit to the last 10 observations (solid line) extrapolated backward (dashed line). Clearly, the first points are strongly inconsistent with a smooth rise.
The date of the NOT spectrum is also indicated.}
\label{fig:LC}
\end{center}
\end{figure*}

Photometric observations were carried out with the SDSS telescope at Apache Point Observatory \citep{1998AJ....116.3040G,2006AJ....131.2332G} in the SDSS $ugriz$ filters \citep{1996AJ....111.1748F}.
As for all SNe discovered by the SDSS SN survey, the identification was made according to \cite{2008AJ....135..348S} and the photometry was performed in the way described in \cite{2008AJ....136.2306H}, to which we refer the reader for more details.
We just point out that these are `\textit{asinh}' magnitudes, defined by \cite{1999AJ....118.1406L} to be identical to the traditional astronomical magnitudes at higher signal-to-noise ratio, but which provide a well-behaved finite flux and error even in the low-flux regime. 
We have derived more traditional 3$\sigma$ upper limits (see Table~\ref{tab:photom}) for non-significant detections (error $\gtrsim$ 0.5 mag), but used the \textit{asinh} magnitudes for the construction of our bolometric light curve (Sect.~\ref{subsec:bolom}).

The SDSS light curves are plotted in Fig.~\ref{fig:LC}. As can be seen,
our observations cover the rise of SN 2006oz, and end before or close to maximum light.
Because of a six day gap in the observations, the explosion date is not very well constrained, but the $g$-band limit previous to our first detection suggests that the explosion occurred between MJD 54022 and 54028.
Nevertheless, these $ugriz$ light curves are very interesting: 
there are hints in our photometry that the SN initially faded in the $u$-band and re-brightened a few days later.
This behavior has only been observed  for a handful of stripped core-collapse SNe that were discovered at very early phases 
and is usually attributed to a phase of adiabatic cooling of the envelope following  shock breakout 
\citep[e.g.][]{1994AJ....107.1022R,2002AJ....124.2100S}. 
Our observations are only poorly constraining (because we are in the low-flux regime), but based on the error bars and upper limits we estimate that there is a probability of 78\% that the SN  faded between MJD 54028 and 54037 in the $u$-band. 
Additionally, a (significant) dip in the light curve is observed simultaneously in the g- and r-bands. 
Combining the $ugr$-bands there is strong evidence ($>$5$\sigma$) for a non-standard, non-increasing luminosity evolution of this SN during the first ten days of observations.

\begin{figure}
\begin{center}
\includegraphics[width=\columnwidth]{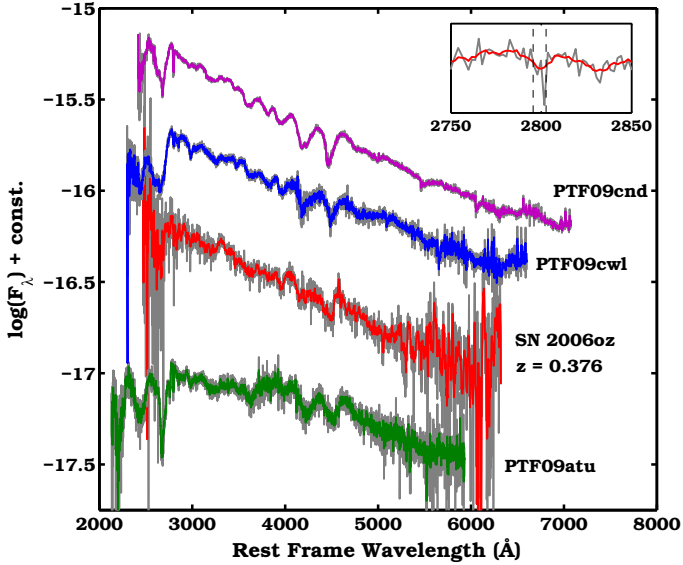}
\caption[]{Spectrum of SN~2006oz (red) obtained at the NOT. For comparison we show 3 spectra from \cite{2011Natur.474..487Q} at similar pre-maximum phases.
The colored spectra are smoothed versions (moving average of 5 pixels) of the original gray spectra. 
All spectra were plotted in their rest frame assuming $z=0.376$ for SN~2006oz. 
The inset shows a zoom of the area around 2800 \AA. 
At this redshift we  identified the only probable line (although at low significance 2.7$\sigma$)
consistent with the \ion{Mg}{ii} doublet (which would appear blended at this resolution).
}
\label{fig:SpecComp}
\end{center}
\end{figure}

At MJD 54061.87, we obtained a spectrum of SN~2006oz at the NOT equipped with ALFOSC.
The spectrum spans the wavelength range 3200--9100 \AA\ with a resolution of 20 \AA\ 
and was reduced in the  manner described by  \cite{2011A&A...526A..28O}. 
The spectrum is displayed in Fig.~\ref{fig:SpecComp} with a sample of comparison spectra from \cite{2011Natur.474..487Q},
obtained at comparable pre-maximum phases,
to highlight their similarity.
To  obtain a redshift estimate, our spectrum was initially cross-correlated with the four objects presented in  \cite{2011Natur.474..487Q} by  using SNID \citep{2007ApJ...666.1024B}. The constructed templates matched reasonably well, all within $0.365 < z< 0.397$ with a mean $z= 0.376 \pm 0.014$. 
It is possible, however, that there are intrinsic velocity differences between the features of SN~2006oz and the other SNe, which would influence our redshift estimate.
We have searched in the wavelength window $\sim$3500-4500~\AA\ for any absorption lines, 
in particular  \ion{Mg}{ii} $\lambda\lambda$2795, 2802  \citep{2011Natur.474..487Q} that would allow us to determine the redshift more accurately.
At the low resolution of our spectrum the doublet will appear blended, making the search more complicated.
The only candidate line that we were able to identify is at $\sim$3856 \AA. Although at low significance ($\sim$2.7$\sigma$), it appears consistently in all sets of extractions that we attempted and always with the same profile (Fig.~\ref{fig:SpecComp}; inset).
Although this line is insufficient  to derive a robust spectroscopic redshift (i.e. \ion{Mg}{ii}  at $z = 0.377$), it can be considered as supporting evidence for the redshift derived by cross-correlating the SN spectra.

Our best redshift estimate, used throughout the paper, is therefore $z= 0.376$.
Assuming a cosmology with $H_0 = 71$ km s$^{-1}$ Mpc$^{-1}$, $\Omega_{\rm{m}} = 0.27$ and  $\Omega_{\Lambda} = 0.73$, SN~2006oz reached an absolute (rest-frame) magnitude of $M_u =-$21.5 mag. 
This calculation contains the assumption that maximum light occurred in our last epoch of observations and is, therefore, only a lower limit to the peak luminosity of the SN. 
In addition, owing to the uncertainty in the cross-correlation redshift, one needs to assign a 0.1 mag systematic error in all quantities that depend on the distance to the SN.
Similarly, we are able to place a strict constraint on the (rest-frame) rise time of $>$29 days. 
Rest-frame wavelengths and related quantities will suffer from an uncertainty of 1\%.

\begin{figure}
\begin{center}
\includegraphics[width=\columnwidth]{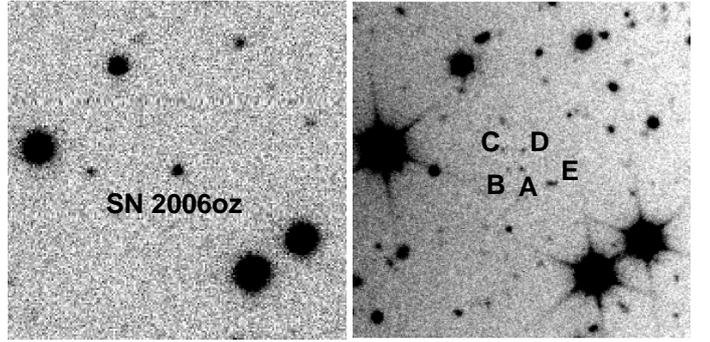}
\caption[]{\textit{Left:} image of SN~2006oz obtained with the SDSS telescope (MJD 54061.17).  \textit{Right:}  deep $i$ image of the field, obtained almost 5 yr after the SN with GTC/OSIRIS. The SN is coincident with the position of galaxy A. However, within a radius of 7\arcsec, there are four more objects, including one that appears extended, labeled B, C, D and E, respectively. Their photometry is reported in Table~\ref{tab:Hostphotom}. Both images are 72\arcsec $\times$ 72\arcsec. North is up and east is to the left.}
\label{fig:Host}
\end{center}
\end{figure}

\begin{table}
\caption[]{Host galaxy candidates for SN~2006oz.}
\label{tab:Hostphotom}
\centering
\begin{tabular}{ccccc}
\hline\hline
id &  $g$ &  $r$ & $i$ &  dist. \\	 
  &  (mag) &  (mag) & (mag) &  (arcsec) \\	 
\hline
A 	&    25.74 $\pm$ 0.19	&  24.43 $\pm$ 0.06 		&  24.14 $\pm$ 0.12	&  0.47 \\	 
B 	&   $>$26.47			&  25.62 $\pm$ 0.16 	 	&  25.02 $\pm$ 0.25	&  2.97\\	 
C 	&  $>$26.47			&  25.81 $\pm$ 0.19 	 	&  24.65 $\pm$ 0.18 	 & 6.26 \\	 
D 	&  25.18 $\pm$ 0.11		&  24.70 $\pm$ 0.07 	 	&  24.32 $\pm$ 0.12	 & 4.56 \\	 
E	&   25.28 $\pm$ 0.15		&  23.71 $\pm$ 0.03 		&  23.21 $\pm$ 0.06 	 & 6.56 \\	 
\hline
\end{tabular} \\
\tablefoot{The indexing follows the notation in Fig.~\ref{fig:Host}. All magnitudes are given in the AB system. 
The last column contains the angular distance from the SN position (as determined by a geometrical transformation between images) and 
has an associated error of 0.35\arcsec.}
\end{table}

On 26 August 2011, almost 5 yr after the SN, we obtained deep $gri$ imaging of the field with OSIRIS on Gran Telescopio Canarias (GTC).
The SN position is spatially coincident with a candidate host galaxy (object A; Fig.~\ref{fig:Host}), but within a radius of 7\arcsec\  there are four more galaxies, including one that appears extended (object E). The photometry of these galaxies and their angular distances from the SN position are presented in Table~\ref{tab:Hostphotom}. The error of these angular distances as determined from the rms of the 
IRAF\footnote{IRAF is distributed by the National Optical Astronomy Observatory: \url{http://iraf.noao.edu/iraf/web/}.} 
task {\tt geomap} was estimated to be 0.35\arcsec. Because of its spatial proximity, we identify galaxy A as the host of SN~2006oz. 
Assuming $z = 0.376$, the SN occurred at  a distance of 2.40 $\pm$ 1.36 kpc from the galaxy center, while the distance to the nearest other candidate would exceed 15 kpc. 
At this redshift, the host of SN~2006oz is intrinsically faint with a (rest-frame) absolute magnitude of $M_g = - 16.9$.
We note that the photometric redshift obtained for this galaxy \citep[using the code \textit{Le Phare};][]{1999MNRAS.310..540A,2006A&A...457..841I} is
 $z_{\rm{phot}} = 0.37^{+3.72}_{-0.04}$. Although not constraining, this value is  consistent with what was obtained for the SN.

\section{Discussion}
\label{sec:disc}

\subsection{Black-body fits}
\label{sec:BBfits}

\begin{figure}
\begin{center}
\includegraphics[width=\columnwidth]{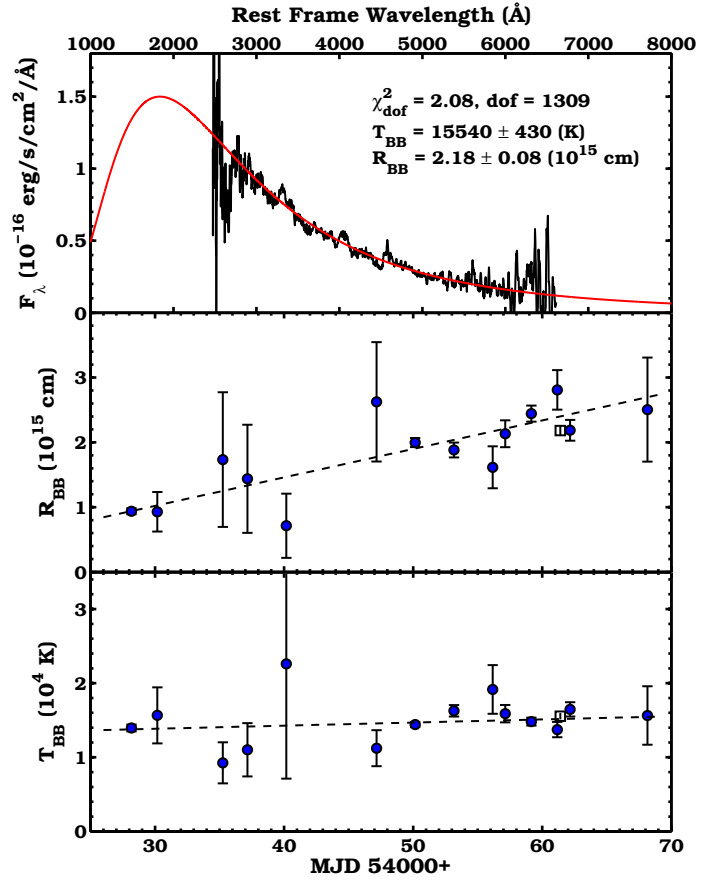}
\caption[]{Black-body fits to the SN~2006oz data. In the top panel, a fit to the NOT spectrum is shown.
Only data redward of  3000~\AA\ (rest-frame) are included in the fit.
The middle and  bottom panel show the radius and temperature evolution of SN~2006oz assuming it can be described by a black body. 
The filled circles are the results of fitting the $griz$ photometry, while the spectrum fit is represented by a square symbol. 
The dashed lines show linear weighted fits to the data.
}
\label{fig:BB3panels}
\end{center}
\end{figure}

To estimate the photospheric temperature and radius of SN~2006oz,  
we fitted a black body (BB) function to the optical spectrum.
As  also pointed out by  \cite{2011ApJ...743..114C},
these events miss some flux in the UV with respect to a BB,
and therefore reliable BB fits can only be obtained by fitting data redward of rest-frame 3000~\AA\ (i.e. the  Rayleigh-Jeans tail).
This was also the experience we gained by fitting the NOT spectrum.
We achieved a reasonable fit with $\chi_{\rm{dof}}^{2} = 2.08$ over 1309 data points, 
resulting in T$_{\rm{BB}} = 15540 \pm 430$ K and  R$_{\rm{BB}} = (2.18 \pm 0.08) \times 10^{15}$~cm.
It is also possible to fit the broad-band photometric data, although with higher uncertainties. 
We therefore performed a weighted fit of the $griz$ data, excluding the $u$-band that corresponds to rest-frame 2574~\AA.
The fit results are shown in Table~\ref{tab:BBfit} and Fig.~\ref{fig:BB3panels}
and  indicate that the emitting radius increased from $\sim$1 to $> 2 \times 10^{15}$~cm during the period of our observations. 
A linear increase describes  the R$_{\rm{BB}}$ evolution well with a weighted fit resulting in $\chi^2_{\rm{dof}}=1.21$.
In contrast, the temperature does not show any significant evolution: a linear fit results in a shallow slope, 
while fitting a constant temperature ($\sim$15000~K) is also consistent with the observations ($\chi^2_{\rm{dof}}=1.92$ versus $1.27$).
However, during the first ten days the average temperature might have been lower ($\sim$12400~K).
The orders of magnitude obtained  and the R$_{\rm{BB}}$ evolution are consistent with what has been  reported previously, although a cooling of the BB is usually observed \citep[e.g.][]{2011Natur.474..487Q,2011ApJ...743..114C}. However, cooling is only observed conclusively at post-maximum phases, while, as discussed below, we are probing earlier stages of SLSN evolution with SN~2006oz. 

\begin{table}
\caption[]{Black-body fits to SN~2006oz.}
\label{tab:BBfit}
\centering
\begin{tabular}{ccc}
\hline\hline
MJD&  T$_{\rm{BB}}$ &  R$_{\rm{BB}}$  \\	 
  &  (K) &  ($10^{15}$ cm)  \\	 
\hline
54028.17 &   13960 $\pm$     460  &    0.94 $\pm$     0.05 \\
54030.18 &   15660 $\pm$    3780  &    0.93 $\pm$     0.30 \\
54032.17 &     $\cdots$           &          $\cdots$      \\
54035.25 &    9253 $\pm$    2775  &    1.73 $\pm$     1.04 \\
54037.16 &   11020 $\pm$    3594  &    1.44 $\pm$     0.83 \\
54040.17 &   22610 $\pm$   15485  &    0.71 $\pm$     0.50 \\
54047.16 &   11230 $\pm$    2425  &    2.62 $\pm$     0.92 \\
54050.16 &   14420 $\pm$     350  &    2.00 $\pm$     0.07 \\
54053.16 &   16270 $\pm$     770  &    1.88 $\pm$     0.12 \\
54056.16 &   19160 $\pm$    3280  &    1.61 $\pm$     0.32 \\
54057.13 &   15890 $\pm$    1170  &    2.13 $\pm$     0.21 \\
54059.16 &   14820 $\pm$     540  &    2.44 $\pm$     0.12 \\
54061.17 &   13740 $\pm$    1030  &    2.81 $\pm$     0.30 \\
54061.87\tablefootmark{a} &   15540 $\pm$     430  &    2.18 $\pm$     0.08 \\
54062.16 &   16470 $\pm$     970  &    2.19 $\pm$     0.16 \\
54068.15 &   15640 $\pm$    3940  &    2.50 $\pm$     0.80 \\	 
\hline
\end{tabular} \\
\tablefoot{
\tablefoottext{a}{This epoch is a fit to the spectrum.}
}
\end{table}

\subsection{Bolometric and absolute light curves}
\label{subsec:bolom}

\begin{figure}
\begin{center}
\includegraphics[width=\columnwidth]{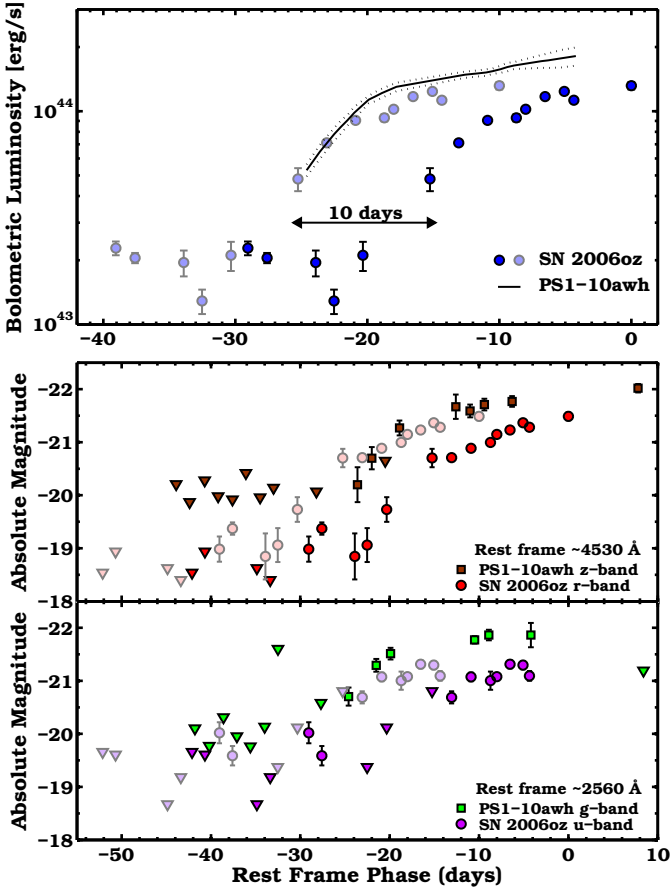}
\caption[]{\textit{Upper panel:} Bolometric light curve of SN~2006oz (circles) compared to  PS1-10awh  \citep[solid line; data from][]{2011ApJ...743..114C}, which has similar pre-maximum multi-color observations.
To illustrate the uncertainty in $t_{\rm{max}}$ of SN~2006oz, its light curve was shifted horizontally with respect to the limiting case, where our last observation was assumed to be taken exactly at maximum light.
Assuming the two events are similar, we obtain a reasonable match of the light curves by shifting $t_{\rm{max}}$ by $-$10 days (faded symbols).
\textit{Middle and lower panels:} The absolute pre-maximum light curves. 
The absolute luminosities at rest-frame $\sim$4530~\AA\  (observed $r$- and $z$-bands)
and $\sim$2560~\AA\ (observed $u$- and $g$-bands) are compared, after correcting for cosmological expansion.
As above, vivid and faded colors show the light curves of SN~2006oz shifted by 10 days.
A few (non-constraining) upper limits were removed for presentation purposes.
The early observations of SN~2006oz are below the detection limits of the PanSTARRS1 observations of PS1-10awh.
}
\label{fig:BolomComp}
\end{center}
\end{figure}

To construct a (pseudo-) bolometric light curve of SN~2006oz we  integrated the flux included within the limits of our simultaneous $ugriz$ observations. 
To account for the epochs containing upper limits (in $u$ and $z$) we  used the \textit{asinh} magnitudes  provided by the SDSS photometry \citep{1999AJ....118.1406L,2008AJ....136.2306H}, e.g. 2.6$\pm$1.5 $\mu$Jy  for our $u$ observation at MJD 54037. This is the most reliable  assumption and the high uncertainties are propagated to the final bolometric light curve. 
A correction was applied only for foreground Galactic extinction \citep[$E(B-V)=0.047$;][]{1998ApJ...500..525S} but not for any host galaxy extinction.
The bolometric light curve is plotted in Fig.~\ref{fig:BolomComp} (upper panel) together with that of PS1-10awh ($z = 0.9084$),
which is the only other  H-poor SLSN with 
a pre-maximum bolometric light curve (based on multi-filter observations) extending back to more than a week before maximum light.
We  constructed the bolometric light curve of PS1-10awh following the same procedure as for SN~2006oz based on the data of \cite{2011ApJ...743..114C}, 
but did not apply a bolometric correction for the red tail of the BB  (this is why our light curve might appear a little different from theirs). 
Determining a bolometric correction is not trivial. 
The rest-frame wavelengths probed were  $\sim$2360-6980~\AA\ (and $\sim$2270-4720~\AA\ for PS1-10awh; extending the integrations out to 0.5 $\times$ FWHM from the central wavelengths of the bracketing filters).
 Assuming the limiting case that these SLSNe can indeed be described by a BB, we can roughly estimate the bolometric correction by measuring the area below the BB curve  covered by our observations. 
For the BB fit to the spectrum of SN~2006oz, the probed wavelength window captures 51\% of the emitted flux, while another 43\% is at bluer wavelengths and only 6\% in the red tail. The corresponding percentages for PS1-10awh are 40\%, 44\%, and 16\%, respectively.
These rough numbers illustrate that the bolometric luminosities obtained could be subject to a correction of up to a factor of 2-2.5.
This is the worst case, however, 
because the flux in the UV clearly deviates from a BB \citep[see also][]{2011ApJ...743..114C}.

In addition to the uncertainty of the total bolometric luminosity,
there is also an uncertainty of the time of maximum. This is illustrated in Fig.~\ref{fig:BolomComp} 
by applying a horizontal shift to the LC  with respect to the limiting case for which we assume that our last observation was exactly at maximum light
(vivid and faded colors). This experiment was made  to show that it cannot be established whether SN~2006oz was a fainter event 
or whether we are probing earlier stages of its evolution (or a combination of both).
Indeed, a reasonable match of the light curves can be obtained by shifting $t_{\rm{max}}$ by $-10$ days (rest-frame, time-dilation corrected).
It is certainly true that the very early data-points of SN~2006oz were below the sensitivity of the PanSTARRS1 observations of  PS1-10awh at $z\sim 0.9$. 
This can also be seen in the lower panel of Fig.~\ref{fig:BolomComp}, which contains a comparison of absolute rest-frame light curves.
In particular, the observed $u$ and $r$-bands of SN~2006oz are compared to the $g$ and $z$ observations of PS1-10awh, which  correspond to the same rest-frame wavelengths.
This comparison shows that (despite the uncertainty in $t_{\rm{max}}$) 
the SDSS observations of SN~2006oz reach deeper limits in an absolute level and provide a window to the early evolution of an SLSN.

The shape of this very early bolometric light curve is very interesting.
The individual band observations discussed in the previous section were translated into a \textit{plateau-like} phase in the 
bolometric light curve. This plateau has a minimum duration of six days and  is followed by a dip, after which the luminosity begins to rise.
Owing to the uncertainty in the  time of explosion, the duration of this initial phase  can be up to 14 days.
Such a precursor plateau has not been observed before for an SLSN, so the reasonable question is whether it is intrinsic to all similar explosions or special to SN~2006oz.  
Its detection was possible by a combination of the SN redshift with multi-color observations including blue filters, a criterion not met by other H-poor SLSNe.
The pre-maximum observations of the PTF SNe \citep{2011Natur.474..487Q} were mostly obtained in the $R$-band and do not extend to the rest-frame UV, while the PS1 observations of \cite{2011ApJ...743..114C} are not as deep, due to the longer distances (Fig.~\ref{fig:BolomComp}).
For SCP06F6 \citep{2009ApJ...690.1358B}, there might be hints for such a plateau in the $z$-band but the photometric errors are too large to allow any firm conclusion.
It is therefore an intriguing possibility that this behavior is intrinsic to all events in this class.
It is unlikely that this result will be changed because this bolometric light curve by construction misses some flux in the UV. Instead, the result will more probably be enhanced because the colors of SN~2006oz evolve from blue to red during this period, which reduces the fraction of missing flux with time. 
The nature of the plateau is  discussed in more detail in Sect.~\ref{subsec:models}.

\subsection{The spectrum}
\label{subsec:spec}

The lack of clear hydrogen features is an important clue to the 
nature of this event and the others in its class \citep{2011Natur.474..487Q}. 
It is difficult to avoid the conclusion that the object is hydrogen-deficient,
although the possibility that some hydrogen is present, but difficult to detect,
cannot be rejected.
The lack of clear signatures of helium is a more ambiguous clue. The
prominent lines of \ion{He}{i} in the optical and NIR arise in transitions
to the $n = 2$ level from higher levels. Ambient conditions in
SN photospheres are typically insufficient to populate these
upper levels, consequently  they must be populated by non-thermal \citep{1987ApJ...317..355H,1991ApJ...383..308L} 
or non-LTE \citep{2011MNRAS.414.2985D} processes.
Relevant ionization 
energies are H - 13.6 eV, \ion{He}{i} - 24.6 eV, \ion{C}{i} - 11.3 eV and \ion{O}{i} - 13.6 eV. 
From the absence of obvious helium lines, it is then much less
obvious that an absence of helium can be firmly deduced.
In addition, the wavelength range of the SN~2006oz spectrum (and most similar events), does not cover 
the most prominent He lines, with the exception of $\lambda$5876 , which is found in a noisy part of the spectrum. 
The question whether He 
might be expected 
requires additional  investigation and better wavelength and temporal coverage.

We have applied the parametrized code {\tt SYNOW} \citep{1999ApJS..121..233H,2003AJ....126.1489B}
to model the spectrum of SN~2006oz. 
Although {\tt SYNOW} is based on simplified physical assumptions, 
 it is very useful for identifying spectral features 
that are strongly Doppler-broadened and sometimes heavily blended.
Because {\tt SYNOW} contains many adjustable parameters, we did not attempt
an automated fitting via chi-squared minimization but instead looked for a
reasonable agreement between the observed and the model spectra by eye, using the minimum
number of different atoms/ions in the envelope. The photospheric expansion velocity 
and the temperature of the underlying blackbody radiation 
were initially set and kept fixed during the search for features.
Adopting $v_{\rm{phot}} = 12000$ km s$^{-1}$ and T$_{\rm{BB}} = 14000$ K resulted
in a good fit to most spectral features.
A power-law atmosphere, where the optical depth
as a function of velocity varies as $(v/v_{\rm{max}})^{-n}$ with $n = 7$, has been assumed for all atoms/ions.
Our tests showed that there are no significant differences between the line profiles obtained by
using either power-law or exponential optical depth profiles for the spectral features of SN~2006oz.
The maximum velocity of the envelope $v_{\rm{max}}$ was set to 40000~km~s$^{-1}$,
but this parameter is only weakly constrained by our spectrum.

\begin{figure}
\begin{center}
\includegraphics[width=\columnwidth]{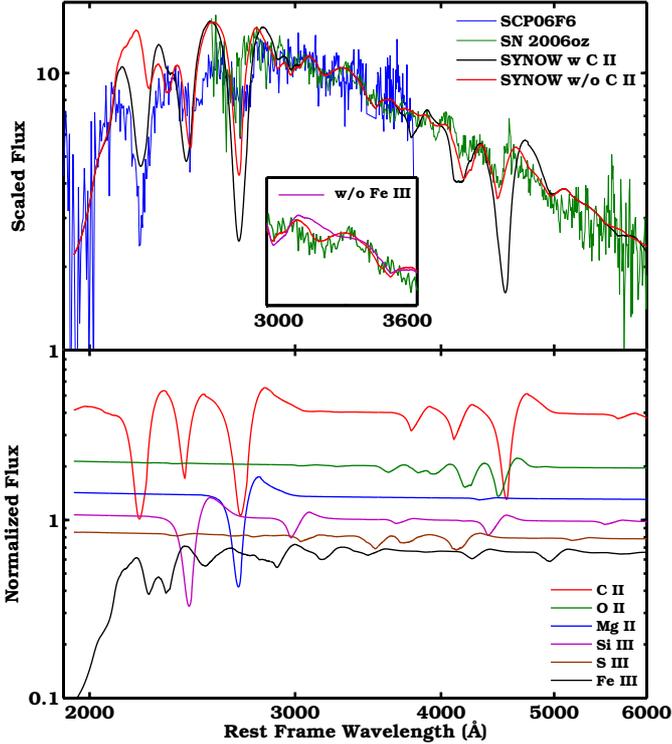}
\caption[]{{\tt SYNOW} fit to the spectra of SN~2006oz and SCP06F6.
The observed spectra were brought to the rest-frame by assuming $z=0.376$ and $z=1.189$, respectively, 
and their fluxes were scaled to match in the overlapping wavelength region. 
Two {\tt SYNOW} models, with and without \ion{C}{ii}, are overplotted. 
The inset shows a zoom in the spectrum of SN~2006oz around 3300 \AA.
The magenta line (inset only) shows a model without \ion{Fe}{iii} that provides a worse fit in this region.
The lower panel shows the contribution of the individual ions to the model spectra. 
The best-fitting model has $\tau$(\ion{O}{ii})~$=$~0.02, $\tau$(\ion{Mg}{ii})~$=$~0.05, $\tau$(\ion{Si}{iii})~$=$~0.5, $\tau$(\ion{S}{iii})~$=$~0.5 
and $\tau$(\ion{Fe}{iii})~$=$~0.4. 
}
\label{fig:SYNOW}
\end{center}
\end{figure}

The observed spectrum and two model spectra, with and without \ion{C}{ii}, are plotted together in
Fig.~\ref{fig:SYNOW}.
The figure also contains the spectrum of SCP06F6 \citep{2009ApJ...690.1358B}
to illustrate a possible extension of the spectrum to the (rest-frame) UV.
The contributions of the individual ions to the model spectrum are also shown.
The ``W''-like feature around 4200~\AA\   nicely matches  the
\ion{O}{ii} feature that was also invoked by \cite{2011Natur.474..487Q} to model other
SLSNe. The strong feature around 2700~\AA\  is
probably due to \ion{Mg}{ii}, although the spectrum is noisy in
that region. \ion{Fe}{iii} can explain the observed bump around 3180~\AA\ (Fig.~\ref{fig:SYNOW}; inset), 
and none of the other ions considered here were able to model that feature.
\ion{Si}{iii} has a contribution around 3000~\AA, but its presence was mostly motivated
by the assumption that SN~2006oz is spectrally similar to other SLSNe.
Indeed, the strongest \ion{Si}{iii} feature matches the observed feature around 2420~\AA\ in
SCP06F6 and in other SLSNe \citep{2011Natur.474..487Q}.
We have also tried to include \ion{S}{iii} in the model of SN~2006oz, but its
spectral features are weak and easily obscured by the noise.
We consider the presence of \ion{Si}{iii} and \ion{S}{iii} as a possibility, but they cannot
be directly identified  in SN~2006oz.

This model is consistent with the findings by
\cite{2011Natur.474..487Q} concerning the presence of \ion{O}{ii}, \ion{Mg}{ii} and \ion{Si}{iii}
(assuming there is indeed some similarity among the spectra of these extreme SNe).
In addition, 
we suggest that \ion{Fe}{iii}  might also be present
and it may also be a good candidate to model the sharp drop of the observed
flux around 2000 \AA\ in SCP06F6.
We did not find the presence of  \ion{C}{ii} necessary  to model SN~2006oz. 
This may be partially because  the strongest feature that was explained with this ion by \cite{2011Natur.474..487Q}
is  beyond the observed spectral range (rest-frame 2200~\AA).
On the other hand, Fig.~\ref{fig:SYNOW} illustrates that
\ion{C}{ii} has strong contributions around
2800, 3800, and 4500~\AA, but these features
do not match the observed spectrum. 
There is some ambiguity between the presence of \ion{C}{ii} and that of \ion{O}{ii} plus \ion{Mg}{ii}, 
but the latter ions better match the observed features than \ion{C}{ii} 
alone. 
We also note that for \ion{C}{ii}, unlike for all other ions for which T$_{\rm{exc}} = $ T$_{\rm{BB}}$ was assumed, 
we had to apply a lower excitation temperature (10000~K), 
because hotter temperatures produced unacceptably strong features.

The chemical composition identified above (\ion{O}{ii}, \ion{Mg}{ii},  and maybe \ion{Fe}{iii}, \ion{Si}{iii} and \ion{S}{iii})
is consistent with the ``carbon-burned'' SN atmosphere considered by \cite{1999ApJS..121..233H}.
Although it is probably premature to conclude that SN~2006oz had such an atmosphere,
it is interesting that together with the lack of H and \ion{He}{i} and maybe \ion{C}{ii}
in the spectrum, all  identified features belong to the ions that are
expected to be the strongest at an excitation temperature of $\sim$14000~K
in an atmosphere containing elements that underwent carbon-burning.

\subsection{Models for the light curve}
\label{subsec:models}

The bolometric light curve can be divided into two parts: 
the initial precursor  plateau and the subsequent smooth monotonic rise. 
As discussed above, it is not clear 
 if the maximum of the light curve was observed, a handicap in 
constraining models.

We note that precursor plateaus, some similar in shape to that of SN~2006oz, were found by
\cite{2011MNRAS.414.2985D} in their models of helium star explosions.  
In their models, the plateau is associated with the shock 
breakout, fireball-cooling phase, before the rise to the peak powered by 
radioactive decay. The plateau arises when the temperature decrease at 
the photosphere slows down, an effect associated with the recombination 
of ejecta layers to their neutral state (primarily He, but also CNO elements 
in the helium-rich progenitor models).  The plateau brightness is determined 
by the amount of energy initially deposited by the shock and the size of 
the progenitor envelope. 
The bolometric light curves of \cite{2011MNRAS.414.2985D} typically have a plateau that lasts about ten days and is about a 
factor of 20 to 30 dimmer than the subsequent peak. 

The physics
explored by the precursor plateaus of \cite{2011MNRAS.414.2985D} may be relevant to SN~2006oz and its kin, but the plateau 
these authors find seems to be too dim to directly correspond to that observed in SN~2006oz. 
In SN~2006oz, this plateau is dimmer than
the subsequent peak by about a factor of 8--10 in luminosity (depending on $t_{\rm{max}}$; Fig.~\ref{fig:BolomComp}). 
Because the plateau is brighter 
than in the models of \cite{2011MNRAS.414.2985D},  the material must
be distributed at larger radii.
The structures in the models of \cite{2011MNRAS.414.2985D} are originally in hydrostatic equilibrium
as dictated by the systematics of stellar evolution. A substantially
larger radius suggests that the plateau we observe in SN~2006oz does
not arise in a stellar envelope, but in a circumstellar medium. 
Consequently, 
the only likely explanation that we have found for the plateau is that it might represent
a recombination wave in a CSM that surrounds the progenitor star. 
A possible case of shock breakout from a H-rich dense CSM was discussed by \cite{2010ApJ...724.1396O}.
For SN~2006oz, 
unlike for SNe~IIP, this plateau cannot be caused by recombination of H, which occurs at much lower temperatures (5000--6000~K).
The derived temperature 
also makes  a He recombination explanation unlikely, because most of the He atoms are singly ionized at
$T > 10000$ K, but double ionization starts above $T > 15000$ K \citep{1999ApJS..121..233H}.
On the other hand, a recombination plateau could be consistent with the transition \ion{O}{iii} to \ion{O}{ii} \citep{1999ApJS..121..233H}.
This idea is also potentially consistent with the detection of \ion{O}{ii} in the spectrum of H-poor SLSNe (Sect.~\ref{subsec:spec}).
Interestingly, a precursor plateau was observed in the simulations of \cite{2010arXiv1009.4353B} for SN~Ia explosions in a C--O CSM,
but, as the authors argue, it was  an artifact of the arbitrary initial conditions.
An early plateau is also present in the bolometric light curve of the 1987A-like SN~2006au \citep{2012A&A...537A.140T}.

\begin{figure}
\begin{center}
\includegraphics[width=\columnwidth]{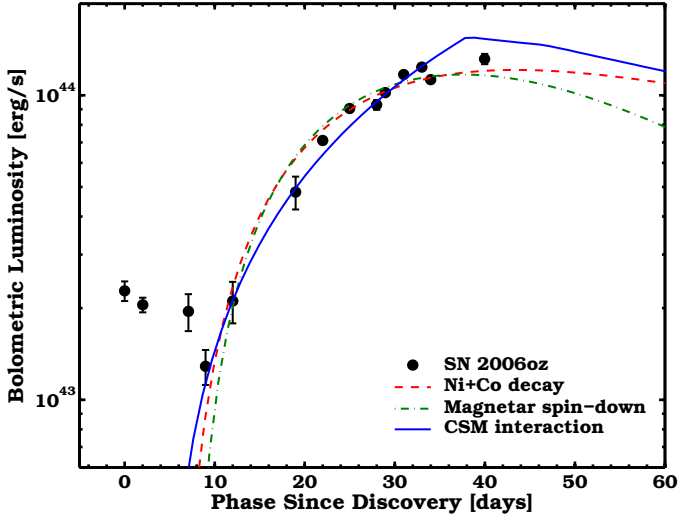}
\caption[]{
Bolometric light curve of SN~2006oz shown along with models describing the rise to maximum (ignoring the precursor plateau).
Models include the following power sources for the luminosity: radioactive decay (red, dashed), input from a magnetar (green, dot-dashed) and CSM interaction (blue, solid).
Details concerning the model parameters are given in the text.
}
\label{fig:model_LC}
\end{center}
\end{figure}

The subsequent smooth rise of the light curve represents a more normal
SN behavior and could be
satisfied by a variety of processes, such as radioactive decay, input from a pulsar or
magnetar, or luminosity from SN ejecta--CSM interaction. 
General semi-analytic models of these processes have
been constructed by \cite{2012ApJ...746..121C}, 
and will be applied to a range of superluminous  events in Chatzopoulos, Wheeler \& Vinko (in prep.). 
Figure~\ref{fig:model_LC} shows a sample of models compared with
the observed bolometric light curve of SN~2006oz, ignoring the
precursor plateau. The models all peak at a time around the
last observed data point, but the lack of a measured maximum 
(and subsequent decline) means the models are not well constrained. 
The models were not fitted in any formal sense, but constructed
to be reasonable ``by eye." The ejecta masses derived may be somewhat
underestimated as discussed by \cite{2012ApJ...746..121C}.

The red dashed curve shows a model driven by radioactive decay with $M_{\rm{Ni}} 
= 10.8$ $M_{\sun}$ and a diffusion time $t_{\rm{d}}$ = 36 days, determined by the 
time scale of the rise of the light curve.  The SN ejecta 
mass, $M_{\rm{ej}}$ = 14.4 $M_{\sun}$, is derived from the diffusion time, $t_{\rm{d}}$, 
assuming a velocity, $v$ = 12000 km s$^{-1}$, and opacity $\kappa = 0.05$~cm$^2$~g$^{-1}$. 
The ejecta mass scales as $\kappa^{-1}$ and consequently  would be lower  for 
higher opacity values. 
The fact that  such a big fraction of $M_{\rm{ej}}$ has been burned to $\rm ^{56}Ni$, unlike in most observed SN types, 
shows that a radioactive decay model  has difficulties  to account for the rise of SN~2006oz to maximum.
Furthermore, the SN was not detected in the next observing season that started nine months later. 
By stacking the images from the first five weeks,
we deduce a limit $r>24.2$ (Table~\ref{tab:photom}) implying a decay rate $>2.03$~mag~$100^{-1}$d$^{-1}$.
This is steeper than the radioactive decay of $\rm ^{56}Co$ ($0.98$~mag~$100^{-1}$d$^{-1}$), assuming complete 
$\gamma$-ray and positron trapping.
The radioactive model for H-poor SLSNe has also been seriously challenged 
by the observations of \cite{2011Natur.474..487Q}, \cite{2010ApJ...724L..16P}, and \cite{2011ApJ...743..114C}.
We stress, though, that an assumption involved in these calculations is that that $\rm ^{56}Ni$ is centrally
condensed  \citep{1982ApJ...253..785A}.
It is therefore possible that a configuration where $\rm ^{56}Ni$ is substantially mixed into the outer envelope, 
thereby allowing $\gamma$-rays to escape freely, cannot be ruled out.

For the same velocity, opacity, and rise time, a magnetar model yields the same ejecta mass, 14.4~$M_{\sun}$, as the radioactive
decay model, but requires two more parameters, $E_{\rm{p}}$ and $t_{\rm{p}}$, where the decay model requires only one, $M_{\rm{Ni}}$, to fit
the rise and peak. 
The green dot-dashed curve shows a magnetar spin-down model with initial rotational energy $E_{\rm{p}} = 1.45\times10^{51}$~erg 
and a characteristic spin-down time of $t_{\rm{p}}$ = 13 days, giving an initial 
rotational period 3.7 ms and a magnetic field $B = 2.24\times10^{14}$~G. 

The blue curve shows a model with SN ejecta and CSM interaction 
(including both forward and reverse shocks)
for again the same velocity, opacity, rise time and ejecta mass as the previous models.
This CSM model was adjusted to give both a reasonable fit to the rise time and a reasonable 
formal black-body temperature at maximum. 
The other parameters used  are \citep[see][for detailed definitions]{2012ApJ...746..121C}:
$E_{\rm{sn}} = 2.5\times10^{51}$~erg, 
the density slope of the SN ejecta $n =12$, 
the progenitor radius $R_{\rm{p}} = 2.5\times10^{14}$~cm,  
the density slope of the CSM $s = 0$, 
the density scale at the base of the CSM $\rho_{\rm{csm,1}} = 5\times10^{-13}$~g~cm$^{-3}$ 
(determined at $R_{\rm{p}}$ and equal to the constant density of the CSM shell), 
$M_{\rm{csm}} = 6.5$~$M_{\sun}$, and
$M_{\rm{Ni}} = 0.02$~$M_{\sun}$. 
The constant CSM photospheric radius of this model
is $R_{\rm{ph}} = 1.8\times10^{15}$~cm, virtually the same as the radius
of the shell itself, giving an optical depth of
$\tau_{\rm{csm}} = 327$. For these parameters and for $L_{\rm{bol}} = 1.3\times10^{44}$~erg~s$^{-1}$ at peak, 
the formal black-body temperature at maximum light  is T$_{\rm BB} = 15300$~K. 
This model, with several free parameters, can also provide a decent fit to the data.

The uncertainty in the bolometric correction and $t_{\rm{max}}$ has, of course, an impact on the models discussed above, although this is not easy to quantify because of parameter degeneracies. 
It is, however, almost certain that constructing a viable radioactive model will become increasingly difficult as 
these uncertainties can only increase the necessary $M_{\rm{Ni}}$ (and the ratio $M_{\rm{Ni}}$ to $M_{\rm{ej}}$) to power the light curve.
Our experience shows that it will still be possible to obtain viable magnetar or CSM models by modifying parameters 
such as $E_{\rm{p}}$, $t_{\rm{p}}$, $\rho_{\rm{csm}}$ and $M_{\rm{csm}}$.
More constraining for these models would be the availability of post-maximum data and decay rate.

We have shown that a range of models can explain the smooth rise to maximum.
Combining these models with the precursor plateau is less straightforward, as
none of these models are presently able to numerically reproduce it.
The possibility that the plateau is the result of a recombination wave in a (O-rich?) CSM, 
the only reasonable suggestion we have come up with, 
gives a rationale for considering that the same shell might be responsible for the rise. 
This picture is not perfectly self-consistent, however, because the first part requires the shock to have broken out of the CSM,
while the second part requires the shock to still be interacting with it.
It is therefore difficult to understand how the CSM was heated in the first place, i.e. before the ejecta--CSM interaction provides the rise to the peak.
A possible answer  could be by energy deposition to the CSM by the SN blast wave, although this possibility remains to be studied.
Another alternative solution could be provided by a hybrid model with a CSM recombination accounting for the plateau and 
a magnetar for the rise to the peak.
Additional  observations of similar events, including more multi-color light curves from very early to very late phases, will be required to resolve this problem.

\subsection{Host Galaxy}
\label{subsec:host}

The host of SN~2006oz is only the fourth out of nine similar H-poor SLSNe 
to have been identified  \citep{2011ApJ...727...15N,2011ApJ...743..114C}.
Very little information is available on the hosts of these events, and many of them are detected in only one filter, 
so that a study of the properties of the host of SN ~2006oz is warranted. 

\begin{figure}
\begin{center}
\includegraphics[width=\columnwidth]{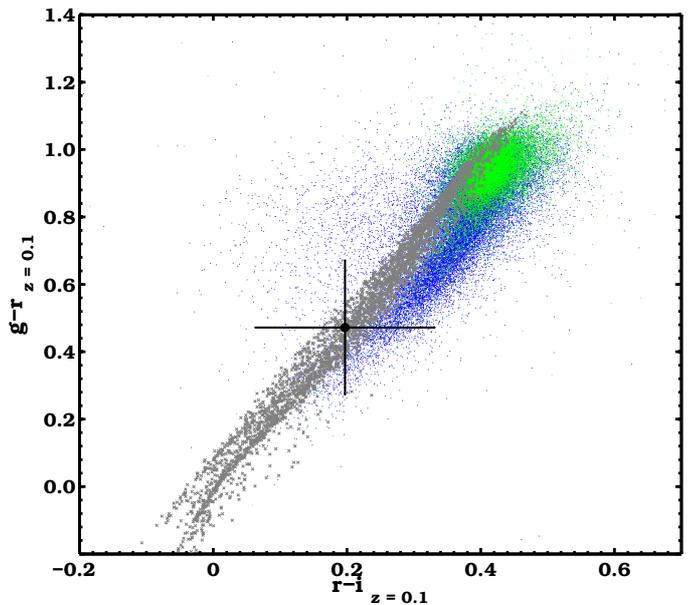}
\caption[]{
Color-color diagram comparing the host of SN~2006oz (black cross) to SDSS galaxies (blue) and to a set of dust-free model galaxies (gray).
The subsample of elliptical galaxies is shown in green.
All data refer (or are $K$-corrected) to $z = 0.1$. 
}
\label{fig:gal_col}
\end{center}
\end{figure}

The colors $g-r=1.26 \pm 0.20$ and $r-i=0.26 \pm 0.13$ (after correcting for Galactic foreground extinction), suggest 
a significant break between the $g$ and $r$ filters. 
This can only be associated with the 4000~\AA\ break, at a redshift consistent with the one we are examining. 
These colors are indeed 
consistent with an intermediate value 
$\rm D_n(4000)=1.4\pm0.1$ for the 4000~\AA\ break,
indicative of relatively old stellar populations and, 
possibly, of a burst of star formation within the last 1--2 Gyr \citep[e.g.][]{2003MNRAS.341...33K,Gallazzi05}.
In Fig.~\ref{fig:gal_col}, the host galaxy colors ($K$-corrected to $z=0.1$)\footnote{The $K$-corrections were estimated from the model that best fits the observer-frame galaxy colors, as described below, and have an uncertainty of 0.13 mag in $g-r$ and 0.03 mag in $r-i$.}
are compared to a set of galaxies, selected\footnote{We  
used the MPA/JHU catalogs available at: \url{http://www.mpa-garching.mpg.de/SDSS/}.}
from the main  spectroscopic sample of SDSS DR4 \citep{2006ApJS..162...38A} to have  $0.08<z<0.12$, 
and to a set of dust-free model galaxies with different star-formation histories \citep{BC03}.
Despite the high uncertainty, the galaxy position on this color-color diagram is more consistent with 
low-mass, star-forming galaxies and is far from the locus of elliptical, passive galaxies.
It is not, however, a blue,  starburst galaxy: 
by using the template of  \cite{1996ApJ...467...38K} and assuming the extinction law of \cite{2000ApJ...533..682C}, 
it is not possible to reproduce these colors for any value of reddening.

To gain a more quantitative understanding of the nature of the host galaxy, its observer-frame colors were compared 
to those predicted by a Monte Carlo library of model  spectra, redshifted to $z=0.376$, based on \cite{BC03}
population synthesis models spanning a wide range of star formation histories, 
metallicities, and dust attenuation \citep[modeled as in][]{2000ApJ...539..718C}. 
In this way, we constructed probability density functions and derived the following estimates for the host properties:
a stellar mass of $\log(M/ M_{\sun}) = 8.7 \pm 0.23$, a luminosity-weighted age of $1.28^{+1.05}_{-0.61}$ Gyr, 
and a SFR of $0.17^{+0.18}_{-0.07}$~$M_{\sun}$~yr$^{-1}$. 
This results in a specific SFR of $\log(\rm{SFR}/M_{\star}) = -9.47 \pm 0.39$ yr$^{-1}$, 
consistent with, but at the lower end of, the values reported by \cite{2011ApJ...727...15N} (although their uncertainties are high).

With an absolute luminosity of $M_g = -16.9$, this galaxy is probably metal-poor. 
By using the mass-metallicity relation of \cite{Tremonti04}, we obtain an indicative value of $12+\log{(\rm{O/H})} = 8.34 \pm 0.16$ for the oxygen abundance that corresponds to 0.45 $Z_{\sun}$ \citep[assuming the solar abundances of][]{2009ARA&A..47..481A}.
A direct metallicity measurement exists only for the host of SN~2010gx \citep{2011ApJ...730...34S}, 
indicating 0.47 $Z_{\sun}$ \citep{2004ApJ...617..240K,2009ARA&A..47..481A}.
These values are not particularly low, but the fact that all H-poor SLSNe have been found in faint galaxies
suggests that metallicity might indeed play a role.
The reason why this is interesting is that all viable models leading to H-poor SLSNe involve 
(one way or another) a very massive star that has suffered extensive mass loss.
Since stellar winds are driven by metals \citep[e.g.][]{vink,2008A&ARv..16..209P} and are less efficient at low metallicity, 
it has been suggested that SLSN progenitors lose mass through episodic outbursts due to mechanisms that are not well understood, possibly involving pulsational pair instability  \citep[e.g.][]{2007Natur.450..390W}.
On the other hand, a similar discussion has been going on about the preference of long GRBs, with optical afterglows, to be found in metal-poor galaxies, but it has been shown that this might be partly due to the fact that smaller galaxies have a higher SFR \citep{2010MNRAS.408.2115M,2011MNRAS.414.1263M}. 
It is therefore possible that the same bias exists for the SLSN hosts.
The fundamental metallicity relation of \cite{2011MNRAS.414.1263M} for low-mass galaxies predicts $12+\log{(\rm{O/H})} = 8.39$ for the host of SN~2006oz, i.e. again a value that is not very low.

\section{Conclusions}
\label{sec:conc}

We have studied SN~2006oz, an event belonging to the family of H-poor SLSNe identified by \cite{2011Natur.474..487Q}. 
The bolometric light curve shows a precursor plateau with a rest-frame duration of 6-10 days,
while the bluest bands demonstrate a dip in the luminosity, before rising smoothly to maximum light.
Our early observations 
are more sensitive than those of previous studies
and it is therefore possible that this behavior might be common to more H-poor SLSN.
If this is the case, 
it provides an important diagnostic for the nature of these events.
We argued that the precursor plateau might be caused by a recombination wave in an H-deficient CSM. 
The subsequent rise in luminosity can be described by energy input from a magnetar or  by ejecta--CSM interaction, 
but radioactive decay is  less likely. 
Although the model parameters are not well constrained owing to the lack of post-maximum data,
all models involve large ejected and CSM masses (if a CSM is required), pointing to a very massive progenitor.

Our deep observations with GTC have revealed the faint  host galaxy of this event ($M_g = -16.9$). 
It is a moderately low-mass ($\log(M/ M_{\sun}) =$ 8.7), 
star-forming (0.17~$M_{\sun}$~yr$^{-1}$) galaxy with a luminosity-weighted age between 0.7-2.3 Gyr. 
It is not a blue, starburst galaxy and the metallicity, inferred indirectly, is not particularly low.

\begin{acknowledgements}

We are grateful to Thomas Kr\"uhler and Lars Mattson for discussions and help.
GL especially thanks Stephen Smartt for the initial encouragement and for providing comments on the manuscript.
GL is supported by the Carlsberg foundation. The Dark Cosmology Centre is funded by the Danish National Research Foundation. 
The research activity of JG and AdUP is supported by Spanish research grants AYA-2011-24780/ESP and AYA2009-14000-C03-01.
JV received support from Hungarian OTKA Grant K76816.
JCW is supported in part by NSF AST-1109801.
JPUF acknowledges support from the ERC-StG grant EGGS-278202.
CCT acknowledges partial funding by project AYA2010-21887-C04-01 ``Estallidos'' of
the Spanish MEC and by FEDER.
We acknowledge the use of the Weizmann Institute of Science Experimental Astrophysics Spectroscopy System. 
Based on observations made with the Gran Telescopio Canarias (GTC), installed in the Spanish Observatorio del Roque de los Muchachos of the Instituto de Astrof\'isica de Canarias, in the island of La Palma.
Funding for the SDSS and SDSS-II has been provided by the Alfred P. Sloan Foundation, the Participating Institutions, the National Science Foundation, the U.S. Department of Energy, the National Aeronautics and Space Administration, the Japanese Monbukagakusho, the Max Planck Society, and the Higher Education Funding Council for England. The SDSS Web Site is \url{http://www.sdss.org/}. The SDSS is managed by the Astrophysical Research Consortium for the Participating Institutions. The Participating Institutions are the American Museum of Natural History, Astrophysical Institute Potsdam, University of Basel, University of Cambridge, Case Western Reserve University, University of Chicago, Drexel University, Fermilab, the Institute for Advanced Study, the Japan Participation Group, Johns Hopkins University, the Joint Institute for Nuclear Astrophysics, the Kavli Institute for Particle Astrophysics and Cosmology, the Korean Scientist Group, the Chinese Academy of Sciences (LAMOST), Los Alamos National Laboratory, the Max-Planck-Institute for Astronomy (MPIA), the Max-Planck-Institute for Astrophysics (MPA), New Mexico State University, Ohio State University, University of Pittsburgh, University of Portsmouth, Princeton University, the United States Naval Observatory, and the University of Washington.

\end{acknowledgements}

%%%%%%%%%%%%%%%%%%%%%%%%%%%%%%%%% REFERENCES %%%%%%%%%%%%%%%%%%%

\bibliographystyle{aa}  %style aa.bst
\bibliography{sn2006oz.bib}

\begin{table*}
\caption[]{SDSS photometry of SN~2006oz (AB magnitudes).}
\label{tab:photom}
\centering
\begin{tabular}{cccccc}
\hline\hline
MJD &  $u$ &   $g$ &  $r$ &  $i$   & $z$   \\	 
\hline
54010.19  &  $>$21.752            &    $>$24.124           &    $>$22.759           &    $>$22.539           &    $>$21.128            \\   
54012.18  &  $>$21.803            &    $>$21.650           &    $>$22.360           &    $>$21.805           &    $>$21.321            \\
54020.18  &  $>$22.740            &    $>$24.200           &    $>$22.673           &    $>$22.688           &    $>$20.332            \\
54022.23  &  $>$22.228            &    $>$23.674           &    $>$22.898           &    $>$22.772           &    $>$20.815            \\
54028.17  &   21.391      (0.198) &    22.280      (0.112) &    22.314      (0.237) &    22.359      (0.388) &    $>$20.666            \\
          &                       &                        &                        &                        &  22.092      (0.856)    \\
54030.18  &   21.823      (0.183) &    21.940      (0.067) &    21.926      (0.116) &    22.268      (0.230) &    $>$21.293            \\
          &                       &                        &                        &                        & 22.683      (0.689)     \\
54032.17  &              $\cdots$ &    21.830      (0.217) &               $\cdots$ &               $\cdots$ &    $\cdots$             \\
54035.25  &   $>$20.899           &    22.591      (0.266) &    22.451      (0.434) &    22.251      (0.425) &    $>$20.184            \\
          &   21.763      (0.440) &                        &                        &                        &  21.152      (0.514)    \\
54037.16  &   $>$21.795           &    22.240      (0.141) &    22.238      (0.320) &    21.858      (0.227) &    $>$21.258            \\
          &   22.861      (0.595) &                        &                        &                        &  23.096      (0.833)    \\
54040.17  &   $>$21.050           &    21.552      (0.216) &    21.567      (0.232) &    22.299      (0.515) &    $>$20.602            \\
          &   22.490      (0.990) &                        &                        &                        &  22.071      (0.904)    \\
54047.16  &   $>$20.364           &    20.924      (0.219) &    20.596      (0.174) &    20.872      (0.304) &    $>$19.567            \\
          &   21.547      (0.713) &                        &                        &                        &  21.306      (1.372)    \\
54050.16  &   20.723      (0.114) &    20.534      (0.032) &    20.590      (0.051) &    20.666      (0.089) &    20.749      (0.378)  \\
54053.16  &   20.337      (0.082) &    20.283      (0.025) &    20.412      (0.040) &    20.507      (0.065) &    20.843      (0.318)  \\
54056.16  &   20.408      (0.172) &    20.160      (0.034) &    20.302      (0.059) &    20.694      (0.145) &    20.571      (0.368)  \\
54057.13  &   20.332      (0.067) &    20.089      (0.021) &    20.150      (0.033) &    20.370      (0.056) &    20.642      (0.225)  \\
54059.16  &   20.097      (0.079) &    20.011      (0.021) &    20.067      (0.035) &    20.177      (0.050) &    20.520      (0.225)  \\
54061.17  &   20.115      (0.079) &    19.964      (0.026) &    19.930      (0.034) &    20.122      (0.052) &    20.098      (0.184)  \\
54062.16  &   20.318      (0.092) &    19.927      (0.030) &    20.017      (0.038) &    20.226      (0.062) &    20.467      (0.285)  \\
54068.15  &              $\cdots$ &    19.800      (0.079) &    19.810      (0.076) &    20.217      (0.161) &    19.909      (0.242)  \\
\hline
54367.02\tablefootmark{a}   &               $\cdots$ &   $>$24.508             &   $>$24.209              &               $\cdots$ &                $\cdots$ \\
54395.57\tablefootmark{b}   &           $>$23.228 &   $\cdots$                  &   $\cdots$                  &           $>$23.592 &             $>$22.190 \\
\hline\hline
\end{tabular} \\
\tablefoot{The quoted limits are 3$\sigma$. For the epochs after the SN discovery, where the SN was below the formal detection limit ($u$ and $z$-band), 
we also provide, below the 3$\sigma$ limits, the \textit{asinh} magnitudes \citep{1999AJ....118.1406L,2008AJ....136.2306H} , which were used as the best approximation in the construction of the bolometric light curve.
\tablefoottext{a}{Observation mid-point and magnitude limits from stacking all observations from the first 5 weeks of the 2007 season.}
\tablefoottext{b}{Observation mid-point and magnitude limits from stacking all observations from the 2007 season.}
}
\end{table*}

\end{document}